\documentclass[useAMS,usenatbib]{mn2e}
\usepackage{graphicx}

\newcommand{\ms}{$\,$M$_\mathrm{\odot}$}

\newcommand{\be}{\begin{equation}}
\newcommand{\ee}{\end{equation}}
\newcommand{\stars}{{\sc stars}}
\newcommand{\el}[2]{\ensuremath{^{#1}\mathrm{#2}}}

\newcommand{\pg}{\ensuremath{\mathrm{(p,\gamma)}}}

\newcommand{\cratio}{\el{12}{C}/\el{13}{C}\ }

\title[Thermohaline mixing in low-Z AGB stars]{The effects of thermohaline mixing on low-metallicity asymptotic giant branch stars}
\author[R.~J. Stancliffe]{Richard J. Stancliffe\thanks{E-mail:
Richard.Stancliffe@sci.monash.edu.au} \\
Centre for Stellar and Planetary Astrophysics, Monash University, VIC 3800, Australia }
\begin{document}
\bibliographystyle{mn2e}

\date{Accepted 0000 December 00. Received 0000 December 00; in original form 0000 October 00}

\pagerange{\pageref{firstpage}--\pageref{lastpage}} \pubyear{0000}

\maketitle

\label{firstpage}

\begin{abstract}
We examine the effects of thermohaline mixing on the composition of the envelopes of low-metallicity asymptotic giant branch (AGB) stars. We have evolved models of 1, 1.5 and 2\ms\ from the pre-main sequence to the end of the thermally pulsing asymptotic giant branch with thermohaline mixing applied throughout the simulations. In agreement with other authors, we find that thermohaline mixing substantially reduces the abundance of \el{3}{He} on the upper part of the red giant branch in our lowest mass model. However,  the small amount of \el{3}{He} that remains is enough to drive thermohaline mixing on the AGB. We find that thermohaline mixing is most efficient in the early thermal pulses and its efficiency drops from pulse to pulse. Nitrogen is not substantially affected by the process, but we do see substantial changes in \el{13}{C}. The \cratio ratio is substantially lowered during the early thermal pulses but the efficacy of the process is seen to diminish rapidly. As the process stops after a few pulses, the \cratio ratio is still able to reach values of $10^3-10^4$, which is inconsistent with the values measured in carbon-enhanced metal-poor stars. We also note a surprising increase in the \el{7}{Li} abundance, with $\log_{10} \epsilon\mathrm(\el{7}{Li})$ reaching values of over 2.5 in the 1.5\ms\ model. It is thus possible to get stars which are both C- and Li-rich at the same time. We compare our models to measurements of carbon and lithium in carbon-enhanced metal-poor stars which have not yet reached the giant branch. These models can simultaneously reproduced the observed C and Li abundances of carbon-enhanced metal-poor turn-off stars that are Li-rich, but the observed nitrogen abundances still cannot be matched.
\end{abstract}


\section{Introduction}

It has long been known that models of asymptotic giant branch (AGB) stars that only include mixing in convective regions are incomplete. These canonical models cannot account for observations such as: the low \cratio ratios in low-mass AGB stars \citep{1997MNRAS.289L..11A, 2008A&A...486..511L}, Li and C-rich stars in our Galaxy \citep{1997MNRAS.289L..11A, 2007A&A...471L..41U}, isotopic ratios measured in pre-solar grains \citep[e.g.][and references therein]{2003ApJ...582.1036N}. It has therefore been suggested that material might circulate below the base of the convective envelope and into regions where nuclear burning can happen. This process is often referred to as `cool bottom processing'. 

The carbon-enhanced metal-poor (CEMP) stars also show abundance trends that are difficult to explain in the context of AGB models including only convective mixing. In particular, those CEMP stars that are rich in $s$-process elements (which we can, with some confidence, assume come from mass transfer in binary systems that once contained an AGB star) show \cratio ratios no greater than 100 \citep[e.g.][]{2006A&A...459..125S}, even for stars at the Main Sequence turn-off. AGB models predict \cratio ratios in excess of $10^4$ \citep{2009MNRAS.396.2313S}. In addition, the models predict that low-mass AGB stars should be rich in carbon, but nitrogen-poor. Yet there is a tendency for these stars to show nitrogen-enhancement alongside their carbon-enhancement\footnote{However, \citet{2009arXiv0901.4737M} note that this trend only holds when considering CEMP stars as a whole. For each of the individual CEMP sub-classes, this trend does not hold.}. Furthermore, there have been detections of lithium in CEMP stars \citep[e.g.][]{2008ApJ...677..556T}. Canonical AGB models produce Li via the Cameron-Fowler mechanism \citep{1971ApJ...164..111C}, which involves the production of beryllium deep in the hydrogen burning shell via the reaction \el{4}{He}(\el{3}{He},$\gamma$)\el{7}{Be} and the immediate transport of this to cooler regions of the star where the \el{7}{Li} that forms (once the beryllium has undergone electron capture) is stable against proton captures. This takes place in the more massive stars which undergo hot bottom burning (HBB, where the base of the convective envelope lies in the top of the hydrogen-burning shell). Such stars would be rich in nitrogen, not carbon. These observations suggest that something is missing from the AGB models and an extra mixing mechanism must be at work.

Lithium is a particularly important element from the point of view of mixing processes in CEMP stars. \citet{2007A&A...464L..57S} pointed out that material accreted on to a low-mass companion does not just remain at the surface of the star and thermohaline mixing could efficiently mix this material deep into the stellar interior. However, detection of lithium in the CEMP binary system CS~22964-161 led \citet{2008ApJ...677..556T} to suggest that the mixing efficiency could not be so high. Li is a fragile element and is easily destroyed at temperatures in excess of about $2.5\times10^6$\,K. Even a modest depth of mixing can lead to efficient Li-depletion \citep{2009MNRAS.394.1051S} and hence the measurement of Li in CEMP stars could be a good test of the efficiency of thermohaline mixing. It is therefore crucial that we understand the origin of this element.

Despite the apparent need for extra mixing on both the giant branches, the physical nature of the mechanism (or mechanisms) has proved illusive. Recently, \citet{2006Sci...314.1580E} showed that the lowering of the mean molecular weight by the reaction \el{3}{He}(\el{3}{He},2p)\el{4}{He} could lead to mixing in red giants via the thermohaline instability. This can potentially explain the change in abundances seen in giants above the luminosity bump \citep{2007A&A...467L..15C, 2008ApJ...677..581E}. The effect of this mechanism has been investigated beyond the first giant branch \citep{2008IAUS..252..103C} and in super-AGB stars \citep{2009A&A...497..463S}. In this work, we wish to examine what the consequences of thermohaline mixing are for low-mass, low-metallicity AGB stars.

\section{The stellar evolution code}
Calculations in this work have been carried out using the \stars\ stellar evolution code which was originally developed by \citet{1971MNRAS.151..351E} and has subsequently been updated by many authors \citep[e.g.][]{1995MNRAS.274..964P,2009MNRAS.396.1699S}. The version used here includes the nucleosynthesis routines of \citet{2005MNRAS.360..375S} and \citet{stancliffe05}, which follow the nucleosynthesis of 40 isotopes from D to \el{32}{S} and important iron group elements. The code uses the opacity routines of \citet{2004MNRAS.348..201E}, which employ interpolation in the OPAL tables \citep{1996ApJ...464..943I} and which account for the variation in opacity as the C and O content of the material varies. An approximation of the contribution to the molecular opacities is included via the method of \citet{2002A&A...387..507M} and is described in \citet{2008MNRAS.389.1828S}\footnote{Recently, low-temperature opacity tables for variable compositions have become available \citep{2009A&A...494..403L,2009arXiv0907.3248M}. These will eventually replace the less accurate approximations used here.}. The modifications for following the evolution through the thermally-pulsing AGB (TP-AGB) phase are described in \citet*{2004MNRAS.352..984S}.

Thermohaline mixing is included throughout all the evolutionary phases via the prescription of \citet*{1980A&A....91..175K}, with the mixing coefficient being multiplied by a factor of 100 as suggested by the work of \citet{2007A&A...467L..15C}. These authors find that with a factor of this magnitude they are able to reproduced the abundance trends observed towards the tip of the red giant branch. \citet{2009MNRAS.396.2313S} also showed that a coefficient of this magnitude could reproduced the observed mixing trends in both low-mass metal-poor stars and carbon-enhanced metal-poor stars on the upper part of the first giant branch.

We evolve stars of 1, 1.5 and 2\ms\ from the pre-main sequence to the end of the thermally pulsing asymptotic giant branch (TP-AGB) using 999 mesh points. \citet{1975MSRSL...8..369R} mass-loss prescription, with $\eta=0.4$, is used from the Main Sequence up to the TP-AGB; the \citet{1993ApJ...413..641V} mass-loss law is employed during the TP-AGB. A mixing length parameter of $\alpha=2.0$ is employed. The metallicity of each model is $Z=10^{-4}$ ([Fe/H]\footnote{[A/B] = $\log_{10}(N_\mathrm{A}/N_\mathrm{B}) - \log_{10}(N_\mathrm{A}/N_\mathrm{B})_\odot$, where $N_i$ is the number abundance of species $i$.}$\approx-2.3$) and the initial abundances are assumed to be solar-scaled according to \citet{1989GeCoA..53..197A}, with the exception of \el{7}{Li}, for which we adopt a value of $\mathrm{X_{^7Li}}=1.05\times10^{-9}$ which is equivalent to the Spite plateau value. It would be more appropriate to adopt an $\alpha$-enhanced composition for these models as the metal-poor stars are observed to have [$\alpha$/Fe]$\approx0.4$. However, opacity tables for an $\alpha$-enhanced mixture that have also been computed with variable C and O abundances do not exist. We therefore choose to have our models self-consistent by keeping our initial compositions solar-scaled like our opacity tables.

\section{Results}

\begin{figure}
\includegraphics[width=\columnwidth]{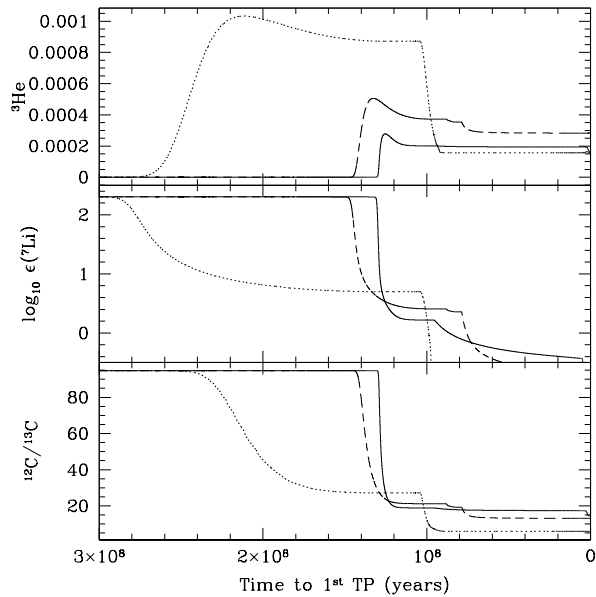}
\caption{Evolution of the surface abundances of various chemical species for the 1\ms\ model (dotted line), 1.5\ms\ model (dashed line) and 2\ms\ model (solid line).}
\label{fig:preAGB}
\end{figure}

The evolution of \el{3}{He}, $\log_{10} \epsilon(\mathrm{^7Li})$ and the \cratio ratio at the surface of each of the models, up to the beginning of the first thermal pulse in each model, is displayed in Fig.~\ref{fig:preAGB}. The behaviour of each model is qualitatively similar. The \el{3}{He} abundance initially increases before dropping slightly and then levelling off. At the same time, the lithium abundance and the \cratio ratio both drop. These changes are all caused by the onset of first dredge-up -- the deepening of the convective envelope as the star ascends the red giant branch. In each of the models, thermohaline mixing begins to affect the surface abundances at around $10^8$ years before the first thermal pulse. The 1\ms\ model is most strongly affected and its surface \el{3}{He} abundance (by mass fraction) falls from around $9\times10^{-4}$ to about $1.5\times10^{-4}$. At the same time, the remaining Li is efficiently destroyed by the mixing and the \cratio ratio drops from its post-dredge-up value of around 30 to around 5.

The two higher mass models bring less \el{3}{He} to the surface during first dredge-up. In addition, their surfaces are less affected by thermohaline mixing on the giant branch. There are only minor reductions in the surface abundances of \el{3}{He}, \el{7}{Li} and the \cratio ratio. The mean molecular weight inversion that formed during the giant branch is not completely removed by the time the star reaches the tip of the giant branch and consequently thermohaline mixing continues to act throughout the core He-burning phase \citep[as was noted by][]{2008IAUS..252..103C}. The surface changes it causes are not dramatic and only the lithium shows a noticeable level of depletion. Eventually, as the convective envelope deepens when the star ascends the early AGB, the surface abundances of \el{3}{He}, \el{7}{Li} and the \cratio ratio all fall slightly as the envelope once again dredges up material that has undergone nuclear burning.

Each of the models is evolved to the onset of the superwind phase at which point numerical problems occur and the runs were terminated. As it is clear that the effects we are interested in only occur at the beginning of the TP-AGB, we need not be concerned with this failure to complete the full thermally pulsing phase. In addition, because each of the models has entered the superwind, mass loss would quickly strip off the remaining envelope and we should not have missed more than a couple of pulse cycles. These missing pulses will not substantially affect the composition of the ejecta. Details of the models are given in Table~\ref{tab:pulses}.

\begin{table*}
\begin{center}
\begin{tabular}{ccccccccccc}
1\ms \\
TP & M$_*$ & $\mathrm{M_H}$ & $\tau_\mathrm{ip}$ & $\log_{10} (L^\mathrm{max}_\mathrm{He}/\mathrm{L_\odot})$ & $\Delta \mathrm{M_H}$ & $\Delta \mathrm{M_{DUP}}$ & $\lambda$ & [C/Fe] & $\log_{10} \epsilon\mathrm{(\el{7}{Li})}$ & \cratio \\
& (\ms) & (\ms) & ($10^4$ yr) & & (\ms) & (\ms)  \\ 
\hline
1 & 0.868 & 0.537 &  ...  & 4.650 & 0.00828 &  ...  &  ...  & -0.67 & ... & 5.7 \\ 
2 & 0.868 & 0.546 & 17.58 & 6.71 & 0.00475 &  ...  &  ...  & -0.67 & ... & 5.7 \\ 
3 & 0.868 & 0.550 & 29.39 & 7.37 & 0.00806 & 0.00017 & 0.021 & 0.61 & 0.32 & 62.9 \\ 
4 & 0.868 & 0.558 & 26.03 & 7.55 & 0.00922 & 0.00142 & 0.154 & 1.80 & 0.55 & 812 \\ 
5 & 0.868 & 0.566 & 17.90 & 7.45 & 0.00884 & 0.00095 & 0.107 & 1.98 & 0.65 & 1164 \\ 
6 & 0.868 & 0.574 & 12.65 & 7.11 & 0.00714 &  ...  &  ...  & 1.98 & 0.72 & 1106 \\ 
7 & 0.867 & 0.581 & 13.91 & 7.52 & 0.00887 & 0.00105 & 0.118 & 2.17 & 0.78 & 1604 \\ 
8 & 0.866 & 0.589 & 12.34 & 7.51 & 0.00864 & 0.00058 & 0.067 & 2.23 & 1.03 & 1649 \\ 
9 & 0.840 & 0.597 & 11.50 & 7.62 & 0.00883 & ... & ... & ... & ... & ... \\ 
1.5\ms \\
TP & M$_*$ & $\mathrm{M_H}$ & $\tau_\mathrm{ip}$ & $\log_{10} (L^\mathrm{max}_\mathrm{He}/\mathrm{L_\odot})$ & $\Delta \mathrm{M_H}$ & $\Delta \mathrm{M_{DUP}}$ & $\lambda$ & [C/Fe] & $\log_{10} \epsilon\mathrm{(\el{7}{Li})}$ & \cratio \\
& (\ms) & (\ms) & ($10^4$ yr) & & (\ms) & (\ms)  \\ 
\hline
1 & 1.436 & 0.586 &  ...  & 5.63 & 0.01152 &  ...  &  ...  & -0.32 & ... & 16.9 \\ 
2 & 1.436 & 0.598 & 12.64 & 7.17 & 0.00552 & 0.00168 & 0.304 & 1.24 & 1.56 & 122 \\ 
3 & 1.436 & 0.601 & 12.05 & 7.45 & 0.00709 & 0.00371 & 0.523 & 1.88 & 2.13 & 350 \\ 
4 & 1.436 & 0.605 & 10.65 & 7.61 & 0.00835 & 0.00544 & 0.651 & 2.25 & 2.33 & 744 \\ 
5 & 1.436 & 0.608 & 10.10 & 7.80 & 0.00959 & 0.00637 & 0.664 & 2.47 & 2.42 & 1229 \\ 
6 & 1.436 & 0.611 & 9.65 & 7.97 & 0.01028 & 0.00686 & 0.667 & 2.63 & 2.46 & 1735 \\ 
7 & 1.435 & 0.614 & 9.16 & 8.05 & 0.01061 & 0.00709 & 0.668 & 2.74 & 2.49 & 2221 \\ 
8 & 1.427 & 0.618 & 8.63 & 8.13 & 0.01068 & 0.00706 & 0.661 & 2.83 & 2.51 & 2672 \\ 
9 & 0.830 & 0.621 & 8.11 & 8.17 & 0.01060 & ... & ... & ... & ... & ... \\ 
2\ms \\
TP & M$_*$ & $\mathrm{M_H}$ & $\tau_\mathrm{ip}$ & $\log_{10} (L^\mathrm{max}_\mathrm{He}/\mathrm{L_\odot})$ & $\Delta \mathrm{M_H}$ & $\Delta \mathrm{M_{DUP}}$ & $\lambda$ & [C/Fe]  & $\log_{10} \epsilon\mathrm{(\el{7}{Li})}$ & \cratio \\
& (\ms) & (\ms) & ($10^4$ yr) & & (\ms) & (\ms)  \\ 
\hline
1 & 1.960 & 0.41469 &  ...  & 5.94 & 0.23905 &  ...  &  ...  & -0.40 & 0.64 & 14.1 \\ 
2 & 1.960 & 0.653 & 6.66 & 7.08 & 0.00429 & 0.00314 & 0.732 & 1.39 & 1.13 & 597 \\ 
3 & 1.960 & 0.654 & 6.59 & 7.50 & 0.00633 & 0.00568 & 0.897 & 1.94 & 1.29 & 2065 \\ 
4 & 1.960 & 0.655 & 6.78 & 7.90 & 0.00823 & 0.00778 & 0.945 & 2.26 & 1.38 & 4219 \\ 
5 & 1.960 & 0.656 & 7.02 & 8.22 & 0.00980 & 0.00924 & 0.943 & 2.46 & 1.43 & 6552 \\ 
6 & 1.960 & 0.656 & 7.13 & 8.45 & 0.01087 & 0.01009 & 0.928 & 2.60 & 1.44 & 8728 \\ 
7 & 1.960 & 0.657 & 7.13 & 8.63 & 0.01157 & 0.01065 & 0.920 & 2.71 & 1.44 & 10684 \\ 
8 & 1.959 & 0.658 & 6.98 & 8.72 & 0.01188 & 0.01079 & 0.908 & 2.79 & 1.43 & 12364 \\ 
9 & 1.956 & 0.659 & 6.77 & 8.78 & 0.01197 & 0.01080 & 0.902 & 2.86 & 1.40 & 13936 \\ 
10 & 1.946 & 0.660 & 6.53 & 8.83 & 0.01195 & 0.01074 & 0.899 & 2.92 & 1.38 & 15278 \\ 
11 & 1.859 & 0.661 & 6.00 & 8.80 & 0.01156 & ... & ... & ... & ... & ... \\ 
\hline
\end{tabular}
\end{center}
\caption{Details of the three models evolved. The columns are: TP -- thermal pulse number; M$_*$ -- total stellar mass in solar masses; M$_\mathrm{H}$ -- hydrogen-exhausted core mass in solar masses; $\tau_\mathrm{ip}$ -- interpulse period in $10^4$ years; logarithm of the maximum helium luminosity reached; $\Delta \mathrm{M_H}$ -- growth of the H-exhausted core mass in solar masses; $\Delta \mathrm{M_{DUP}}$ -- mass of material dredged-up; $\lambda$ -- dredge-up efficiency (which is defined as $\Delta \mathrm{M_{DUP}}/\Delta \mathrm{M_H}$). The remaining columns give the abundances of carbon, lithium and the \cratio ratio measured at the end of the interpulse phase following the thermal pulse. Each model run encountered numerical problems at the peak of a thermal pulse and hence no information for $\Delta \mathrm{M_{DUP}}$, $\lambda$ and abundances could be given for the last pulse.}
\label{tab:pulses}
\end{table*}

\begin{figure}
\includegraphics[width=\columnwidth]{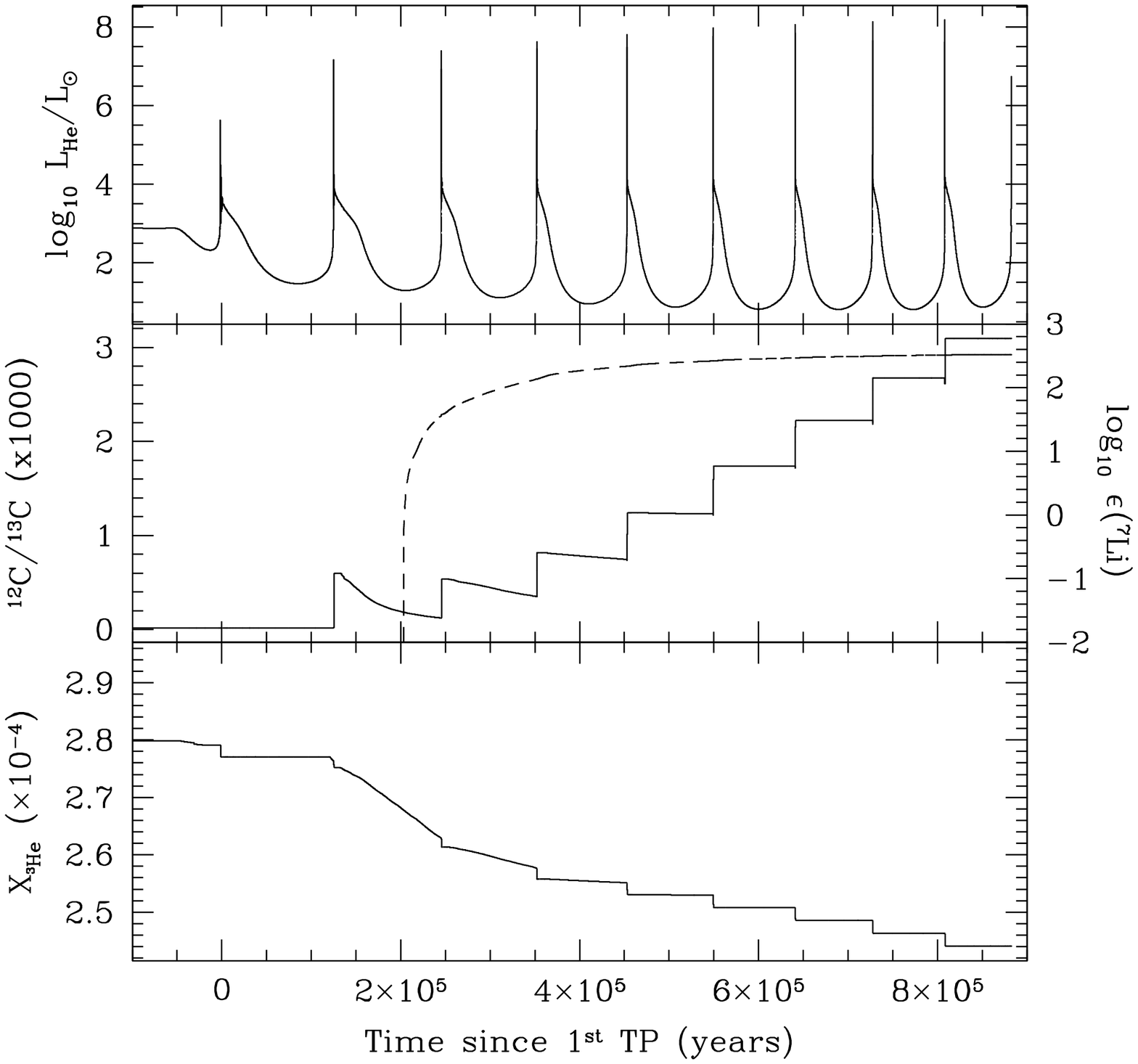}
\caption{Properties of the 1.5\ms\ model as a function of time since the first thermal pulse. Top panel: The evolution of the helium-burning luminosity. Middle panel: The evolution of the surface \el{12}{C}/\el{13}{C} ratio (solid line) and the surface Li abundance (dashed line). Bottom panel: The evolution of the surface \el{3}{He} abundance.}
\label{fig:PulseAbundance}
\end{figure}

The occurrence of mixing by thermohaline convection driven by the burning of \el{3}{He} is seen to occur in each of the models. Mixing is able to affect the abundances of \el{3}{He}, \el{7}{Li} and \el{13}{C}. Heavier elements are not affected as their proton-burning reactions are activated at much higher temperatures than the mixing mechanism is able to reach. We discuss in detail the case of the 1.5\ms\ model, the salient properties of which are displayed in Fig.~\ref{fig:PulseAbundance}. The other models behave in much the same way (see section~\ref{sec:massdiff} for a discussion of the differences between them). The mixing is at its most efficient in the earliest pulses. After the first thermal pulse with dredge-up, the \cratio ratio falls dramatically in the interpulse. Just after the end of third dredge-up (TDUP) the \cratio ratio is around 600 while by the end of the interpulse it has fallen to just 120. At the same time the \el{3}{He} abundances drops from $2.90\times10^{-4}$ to $2.76\times10^{-4}$ and the \el{7}{Li} abundances increases from zero to $1.99\times10^{-10}$  (equivalent to $\log_{10} \epsilon\mathrm{(\el{7}{Li})} = 1.56$\footnote{$\log_{10} \epsilon\mathrm{(X)} = \log_{10}\epsilon(N_\mathrm{X}/N_\mathrm{H}) + 12$, where $N_\mathrm{X}$ is the number abundance of species X and $N_\mathrm{H}$ is the number abundance of hydrogen.}). On the next pulse, the effect is diminished with the \cratio ratio being 540 at the end of TDUP and by the end of the interpulse it has dropped to 350. The \el{3}{He} abundances falls to $2.58\times10^{-4}$ and $\log_{10} \epsilon\mathrm{(\el{7}{Li})}$ reaches 2.14.

Why does the effect fall off so rapidly? The point at which the $\mu$-minimum is located is critical. What determines this is the competition between two things: the rate at which the mean molecular weight can be lowered via \el{3}{He}(\el{3}{He},2p)\el{4}{He} and the rate at which it is increased by other reactions producing \el{4}{He}. In the absence of efficient CNO cycling in the H-shell the $\mu$-minimum can lay deeper in the star as the pp-chains dominate. When dredge-up happens, the enhanced C-abundance increases the efficiency of the CNO cycle. This forces the $\mu$-minimum outward to cooler temperatures and higher \el{3}{He} abundances, with the magnitude of the $\mu$-dip being reduced. The effect of this is two-fold. First, the reduction in the size of the $\mu$-dip makes the mixing less efficient, as the $\mu$-gradient (which the mixing rate is proportional to) is also reduced. As the location of the $\mu$-dip occurs at progressively higher \el{3}{He} abundances from pulse to pulse (the \el{3}{He} abundance at the $\mu$-minimum is around $6\times10^{-5}$ during the second interpulse, $1.5\times10^{-4}$ during the third interpulse and $2\times10^{-4}$ during the fourth interpulse), the rate of \el{3}{He} depletion falls off because material that is less depleted in \el{3}{He} is now being mixed into the envelope. Secondly, the $\mu$-minimum is pushed out to a region of cooler temperature. In this region, the \el{12}{C}\pg\el{13}{N} reaction is less  efficient and the material that is mixed to the surface has a smaller fraction of \el{13}{C} (formed from the decay of \el{13}{N}). This, coupled with the reduction in the mixing efficiency, means that the \cratio ratio falls more slowly with each subsequent pulse.

\subsection{Lithium production}

\begin{figure*}
\includegraphics[angle=270,width=2\columnwidth]{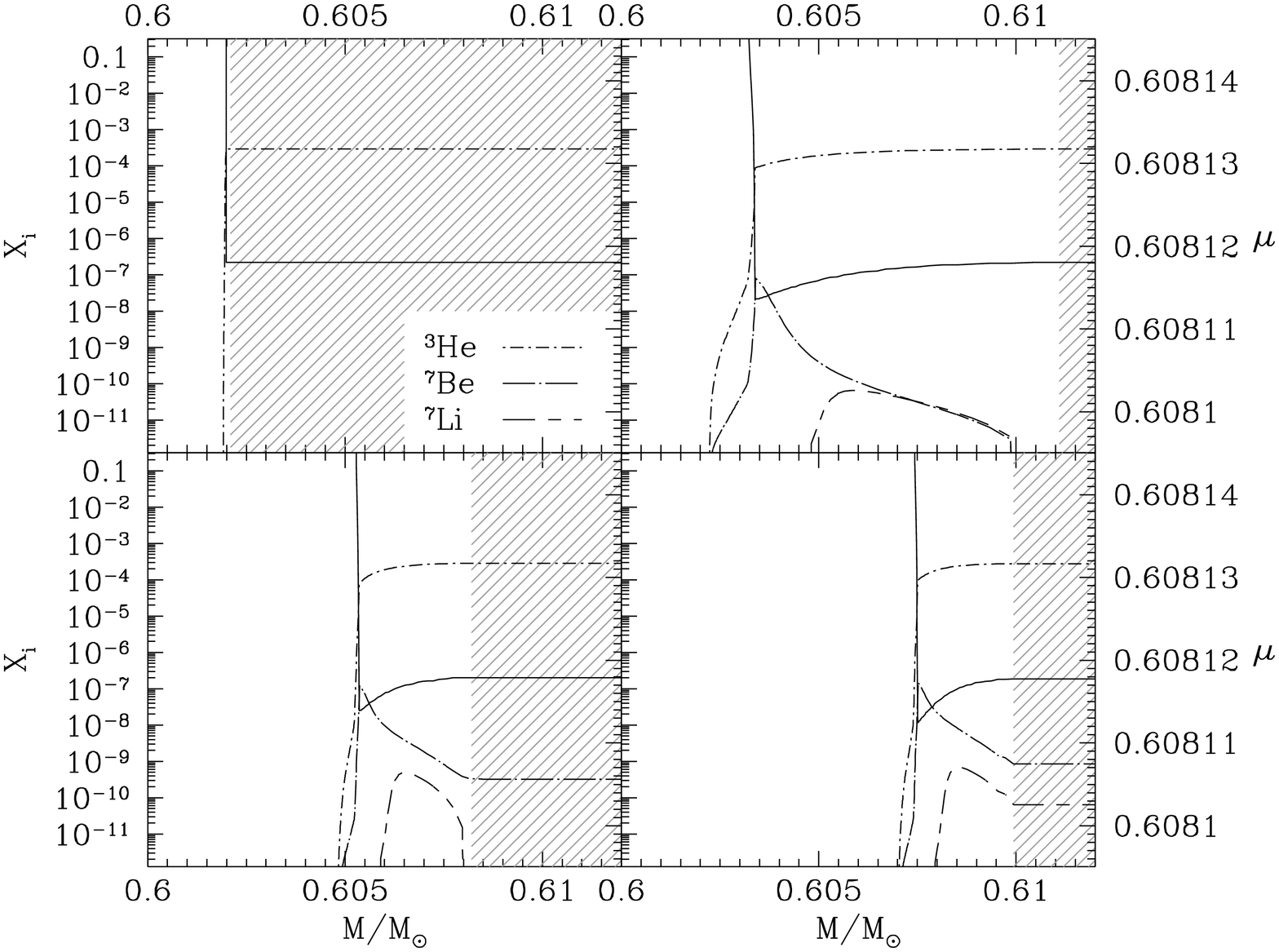}
\caption{Abundance profiles of the 1.5\ms\ model after its second thermal pulse. In each panel, the abundances are: \el{3}{He} (dot-short dash line), \el{7}{Be} (dot-long dash line) and \el{7}{Li} (dashed line). The mean molecular weight, $\mu$ is displayed by a solid line. The grey shading indicates convective regions. {\bf Top left:} Just after the end of third dredge-up. {\bf Top right:} Retreat of the convective envelope. Thermohaline mixing leads to the formation of a pocket of beryllium and lithium. {\bf Bottom left:}. The hydrogen shell moves outward and the envelope comes back in again. {\bf Bottom right} Just before the beginning of the next thermal pulse.}
\label{fig:1.5MsolProfiles}
\end{figure*}

Li production takes place in three stages and begins with the deepening of the convective envelope at TDUP. This homogenises the convective envelope, resulting in a constant mean molecular weight above the soon to ignite H-burning shell. Without this flattening of the $\mu$-profile, the burning of \el{3}{He} would not be able to produce a large enough change in the mean molecular weight to allow thermohaline mixing to take place. After TDUP has occurred, the convective envelope retreats, hydrogen burning reignites and the $\el{3}{He} + \el{4}{He}$ reaction leads to the production of \el{7}{Be} (top panels of Fig.~\ref{fig:1.5MsolProfiles}). At the same time, the \el{3}{He}$(\el{3}{He},2\mathrm{p})$\el{4}{He} reaction leads to a reduction of the mean molecular weight. Once a $\mu$-inversion develops, thermohaline mixing can take place and the beryllium is brought up to regions of cooler temperature where the \el{7}{Li} that is formed from the beryllium is able to survive for longer than it can closer to the burning shell (top right panel of Fig.~\ref{fig:1.5MsolProfiles}). The profiles of \el{7}{Be} and \el{7}{Li} are due to the equilibrium between the rate at which these isotopes are transported up from the \el{3}{He} burning regions (by the action of thermohaline mixing) and the rate at which they are destroyed. A pocket of \el{7}{Li} is formed. As the $\mu$-inversion reduces, the efficiency of mixing is reduced and \el{7}{Be} is no longer mixed outwards as far. All this takes place very shortly after the cessation of TDUP. At no point is the efficiency of the mixing enough to get the Li into the convective envelope before it is destroyed. 

After the H-shell re-ignites the convective envelope moves back inward in mass (bottom left panel of Fig.~\ref{fig:1.5MsolProfiles}) and connects with the upper edge of the pocket of beryllium and lithium. This inward motion of the convective envelope takes place during the interpulse period and it is at this point that the Li enhancement is first seen. The envelope does not move in to as great a depth as it did during TDUP but it brings the base of the envelope close enough to the hydrogen-burning shell that Li can be transported up into the envelope before it is destroyed (bottom right panel of Fig~\ref{fig:1.5MsolProfiles}).

\subsection{Carbon-13}

\begin{figure*}
\includegraphics[angle=270,width=2\columnwidth]{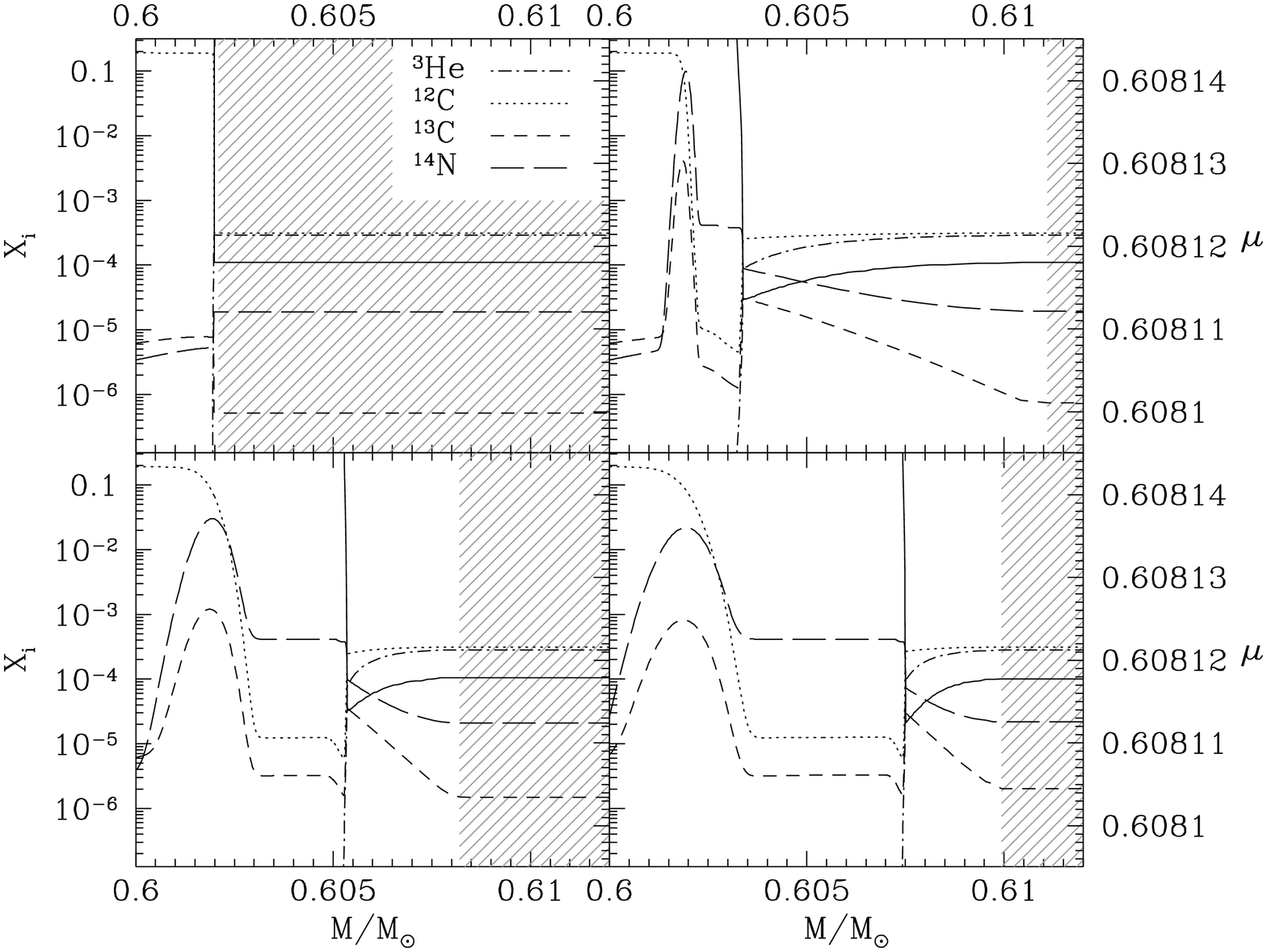}
\caption{Abundance profiles of the 1.5\ms\ model after its second thermal pulse. In each panel, the abundances are: \el{3}{He} (dot-short dash line), \el{12}{C} (dotted line), \el{13}{C} (short-dashed line) and \el{14}{N} (long-dashed line). The mean molecular weight, $\mu$ is displayed by a solid line. The grey shading indicates convective regions. {\bf Top left:} Just after the end of third dredge-up. {\bf Top right:} Retreat of the convective envelope. {\bf Bottom left:} The hydrogen shell moves outward and the envelope comes back in again. {\bf Bottom right} Just before the beginning of the next thermal pulse.}
\label{fig:CNOProfilesTP2}
\end{figure*}

The production of \el{13}{C} is much more straightforward. Abundance profiles for the 1.5\ms\ model during its second interpulse phase are shown in Fig.~\ref{fig:CNOProfilesTP2}, with the isotopes \el{3}{He}, \el{12}{C}, \el{13}{C} and \el{14}{N} being displayed. The newly-dredged up \el{12}{C} is converted into \el{13}{C}, which is then carried to the convective envelope by the thermohaline mixing. Once lifted from the H-burning shell, \el{13}{C} undergoes no further reactions and so we see a steady increase in this isotope throughout the interpulse (unlike \el{7}{Li}, which still undergoes proton captures as it is being transported outward). Thus the \cratio ratio drops throughout the interpulse. There is a concomitant slight increase in the surface \el{14}{N} abundance as the temperature down to which mixing occurs is sufficiently high for incomplete CN-cycling.  At the top of the H-burning shell, nitrogen is enhanced to a few times its surface value after third dredge-up, whereas \el{13}{C} is over an order of magnitude more abundant at the top of the H-shell than it is at the surface. Hence the N-enhancement caused by thermohaline mixing is insignificant (and becomes even less significant with each pulse).

\begin{figure*}
\includegraphics[angle=270,width=2\columnwidth]{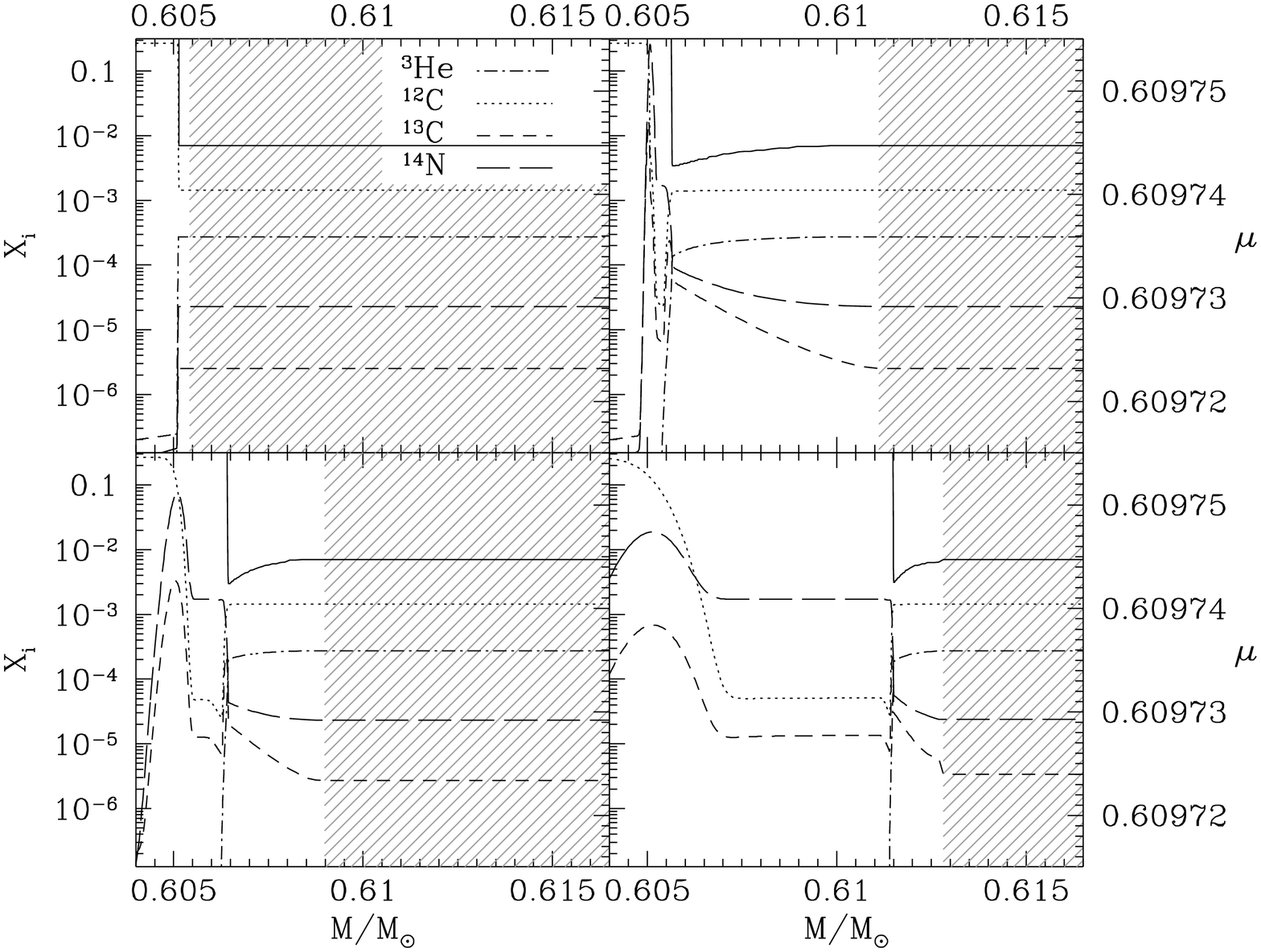}
\caption{Abundance profiles of the 1.5\ms\ model after its second thermal pulse. In each panel, the abundances are: \el{3}{He} (dot-short dash line), \el{12}{C} (dotted line), \el{13}{C} (short-dashed line) and \el{14}{N} (long-dashed line). The mean molecular weight, $\mu$ is displayed by a solid line. The grey shading indicates convective regions. {\bf Top left:} Just after the end of third dredge-up. {\bf Top right:} Retreat of the convective envelope. {\bf Bottom left:} The hydrogen shell moves outward and the envelope comes back in again. {\bf Bottom right} Just before the beginning of the next thermal pulse.}
\label{fig:CNOProfilesTP3}
\end{figure*}

During the third interpulse phase (see Fig.~\ref{fig:CNOProfilesTP3}), the \cratio declines by much less. The dredge-up of carbon forces the $\mu$-minimum out to cooler temperatures. As a consequence of this, the CN cycle is less efficient and less \el{13}{C} is produced: while the \el{12}{C} abundance in the envelope has increased from $3.1\times10^{-4}$ to $1.3\times10^{-3}$, the \el{13}{C} abundance at the $\mu$-minimum is roughly constant at around $3\times10^{-5}$. In addition, the magnitude of the mixing coefficient is about a factor of 2 lower in the third interpulse than it was in the second, owing to the much smaller $\mu$-dip. To compound all this, the interpulse is also $2\times10^4$\,yrs shorter than the preceding one. All this results in the \cratio ratio suffering less of a reduction during the interpulse.

The reduction in the efficacy of mixing continues from pulse to pulse and very quickly thermohaline mixing is no longer able to affect the \cratio ratio. After three pulses there is barely a discernible change in it during the interpulse. The \cratio ratio keeps increasing with each episode of third dredge-up and values of over $10^3$ are reached by the end of the TP-AGB.

\subsection{Mass variations}\label{sec:massdiff}

\begin{figure}
\includegraphics[width=\columnwidth]{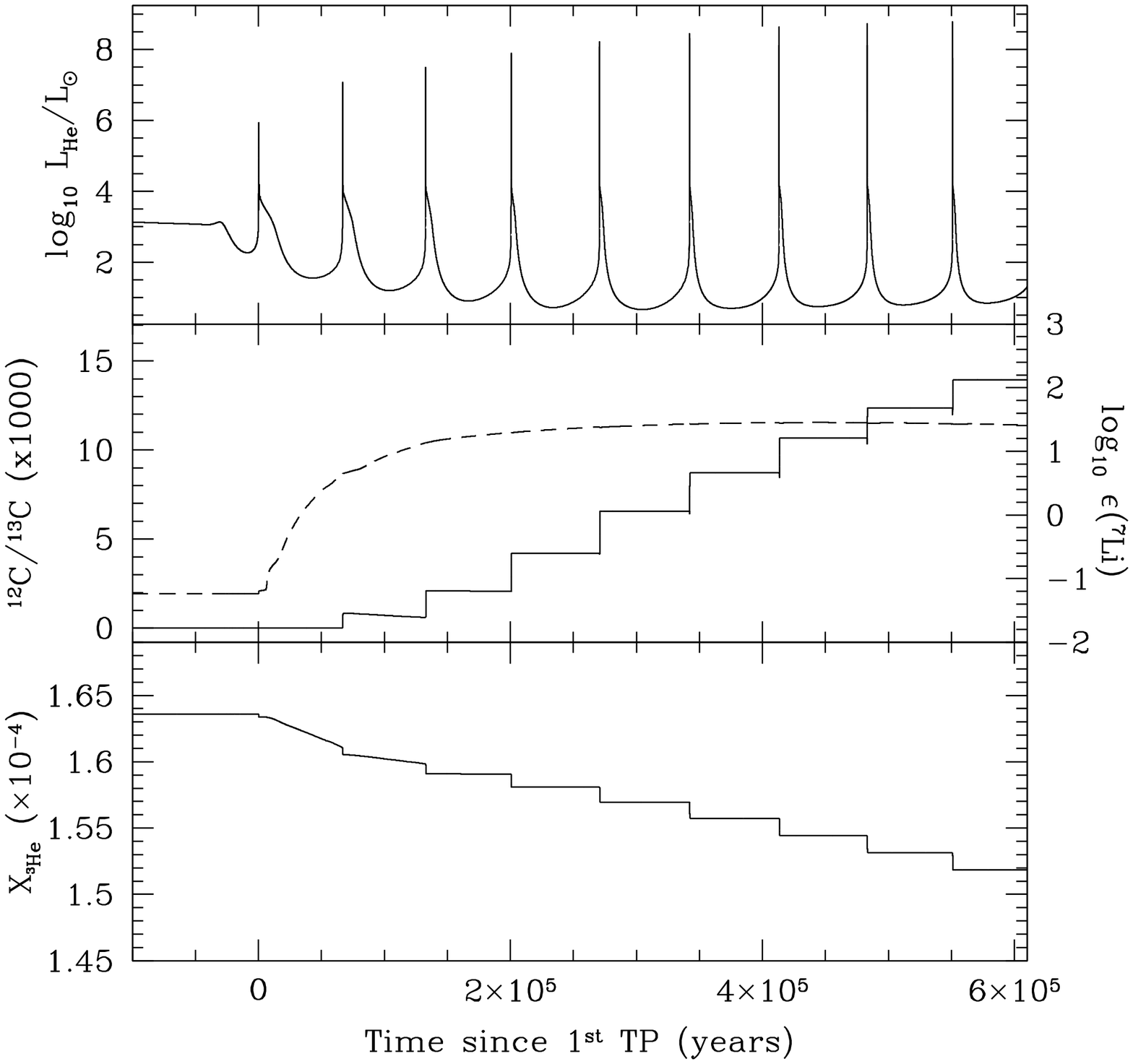}
\caption{Properties of the 2\ms\ model as a function of time since the first thermal pulse. Top panel: The evolution of the helium-burning luminosity. Middle panel: The evolution of the surface \el{12}{C}/\el{13}{C} ratio (solid line) and the surface Li abundance (dashed line). Bottom panel: The evolution of the surface \el{3}{He} abundance.}
\label{fig:2MsolPulseAbundances}
\end{figure}

The 2\ms\ model shows less of an effect on its surface abundances than the 1.5\ms\ model (see Fig.~\ref{fig:2MsolPulseAbundances}). For the \cratio ratio, there is less time for circulation of material close to the H-burning shell because the interpulse periods are considerably shorter (by about a factor of 2) than in the 1.5\ms\ model. In addition, the lower \el{3}{He} abundance that this model starts the TP-AGB with also reduces the efficiency of mixing. The peak Li abundance that is reached is also lower than in the 1.5\ms\ model due to the reduced availability of \el{3}{He} and the reduced mixing efficiency. In addition, we also see a decrease in the Li abundance towards the end of the TP-AGB because the temperature at the base of the convective envelope increases sufficiently to allow Li destruction. We also note that the Li abundance increases directly after the first thermal pulse, even though there is no third dredge-up. This occurs because the convective envelope still moves inward after this thermal pulse but does not reach  deep enough to penetrate the C-rich layers of the intershell. The envelope reaches to the H-burning shell and erases the existing mean molecular weight gradient. It is this flattening of the mean molecular weight gradient by the convective envelope that allows the \el{3}{He} burning to form a mean molecular weight inversion and thus allows thermohaline mixing to occur.

The 1\ms\ model also shows less of an effect on its surface abundances than the 1.5\ms\ model. Again, we attribute this to the lower \el{3}{He} abundance (the model starts the TP-AGB with $X_{\el{3}{He}} = 1.39\times10^{-4}$) which reduces the mixing efficiency. Unlike the 2\ms\ model, the first efficient episode of thermohaline mixing happens after the first episode of third dredge-up. We thus get a substantial reduction of the \cratio ratio during the next interpulse phase.

\section{Uncertainties}

There are many uncertainties associated with thermohaline mixing as a mechanism, most notable of which is the efficiency of the mixing. The prescriptions of \citet{1972ApJ...172..165U} and \citet{1980A&A....91..175K} give diffusion coefficients of the same functional form, but with a dimensionless constant (related to the geometry of the mixing) that differs by orders of magnitude \citep[see][for details]{2007A&A...467L..15C}. There are also differences in the way that one applies the mixing. \citet{2008ApJ...677..581E} use a mixing coefficient that depends on the difference between $\mu$ at a given point in the star and the minimum value of $\mu$. This is a non-local prescription, whereas \citet{2007A&A...464L..57S} use a local prescription where the mixing efficiency depends on local differences in $\mu$ (i.e. the $\mu$-gradient). \citet{2008ApJ...679.1541D} also include overshooting in their prescription so that they actually mix to {\it below} the location of their $\mu$-minimum.

In light of these uncertainties, we make the following tests on our 1.5\ms\ model to determine the robustness of our results. We have run two model sets in which we change the diffusion coefficient of the existing prescription. In one, we increase the diffusion coefficient by a factor of 5\footnote{An increase of a factor of 10 proved to be unstable and the model could not be converged.} (model D5)  and in the other we decrease it by a factor of 10 (model D0.1). We also implement a prescription similar to that of \citet{2008ApJ...677..581E} (hereafter EDL08), where the diffusion coefficient is given by
\be
D_\mathrm{th} = C(\mu - \mu_\mathrm{min}) 
\ee
for regions above the point of minimum $\mu$. Here, $D_\mathrm{th}$ is the diffusion coefficient, C is a free parameter, $\mu$ is the mean molecular weight and $\mu_\mathrm{min}$ is the minimum mean molecular weight in the model. The free parameter C is set to $10^{14}$ as this gives approximately the same carbon depletion on the upper giant branch as is obtained in our standard model. Using this prescription, the mixing is made global rather than local (model EDL). We note that the EDL08 prescription gives a very different functional form for the mixing coefficient to our standard implementation. The EDL08 prescription gives faster mixing {\it further} away from the $\mu$-minimum because this is where $\mu-\mu_\mathrm{min}$ is greatest, whereas our standard prescription gives most efficient mixing close to the point of $\mu$-minimum because this is where the $\mu$-gradient is at its steepest. 

Each of these models is run from the end of first dredge-up until the superwind phase of the TP-AGB. Table~\ref{tab:uncertainties} shows how the \el{3}{He} \& \el{7}{Li} surface abundances and the \cratio ratio at various evolutionary stages change with the varying of the parameters.

\begin{table}
\begin{center}
\begin{tabular}{l|ccc}
\hline
Model & \el{3}{He} & $\log_{10} \epsilon\mathrm{(^7Li)}$ & \cratio \\
\hline
Post $1^\mathrm{st}$ \\
 dredge-up & $3.52\times10^{-4}$ & 0.41 & 21.2 \\ 
\hline
Tip of the \\
giant branch \\
\hline
Standard & $3.27\times10^{-4}$ & 0.31 & 18.6 \\
D5 & $1.95\times10^{-4}$ & -0.30 & 10.1 \\
D0.1 & $3.33\times10^{-4}$ & 0.35 & 19.3 \\
EDL & $3.21\times10^{-4}$ & 0.32 & 17.6 \\
\hline
First thermal pulse \\
\hline
Standard & $2.76\times10^{-4}$ & -1.40 & 11.7 \\
D5 & $1.45\times10^{-4}$ & -1.82 & 8.11 \\
D0.1 & $3.03\times10^{-4}$ & -0.48 & 15.3 \\
EDL & $1.04\times10^{-4}$ & -2.57 & 6.02 \\
\hline
Final model\\
\hline
Standard & $2.44\times10^{-4}$ & 2.51 & 2672 \\
D5 & $1.22\times10^{-4}$ & 2.89 & 2317 \\
D0.1 & $2.92\times10^{-4}$ & 0.61 & 10100 \\
EDL & $9.75\times10^{-5}$ & 0.24 & 10420  \\
\hline
\end{tabular}
\end{center}
\caption{Surface abundances by mass fraction of isotopes affected by thermohaline mixing and how they vary with parameters in the 1.5\ms\ model, together with the \cratio ratio, at various evolutionary stages. The models are all started from the same point after the end of first dredge-up.}
\label{tab:uncertainties}
\end{table}

If we increase the diffusion coefficient by a factor of 5, we get substantially more processing of material on the first giant branch. The drop in the \el{3}{He} during the mixing episode on the upper giant branch is about 6 times as great as for the standard diffusion coefficient. The lithium abundance drops from $\log_{10} \epsilon\mathrm{(^7Li)}=0.41$ to $\log_{10} \epsilon\mathrm{(^7Li)}=-0.3$ and the \cratio ratio drops to just 10.1, compared to 21.2 in the standard case. Despite the depletion in the helium-3 reservoir, the model is still able to undergo thermohaline mixing on the AGB. The total Li production is greater than in the standard case by over a factor of 2, with the final model reaching a surface abundance of $\log_{10} \epsilon\mathrm{(^7Li)}=2.89$. The final \cratio ratio reached is 2317, which is only slightly lower than the standard case. This is because thermohaline mixing is only effective at reducing this ratio in the early pulses and this does not change with changes in the diffusion coefficient.

Reducing the diffusion coefficient by an order of magnitude does not substantially affect the amount of processing that takes place on the first giant branch. The surface abundances of \el{3}{He}, \el{7}{Li} and the \cratio ratio are almost identical in the two models. The slower diffusion coefficient means there is less of a reduction in the \el{3}{He}, \el{7}{Li} and \cratio ratio between the tip of the giant branch and the beginning of the TP-AGB. Along the TP-AGB, the reduction of the diffusion coefficient shows a much more significant effect. The reduction in the efficiency of mixing means that less \el{7}{Li} is brought into the envelope and the model reaches only $\log_{10} \epsilon\mathrm{(^7Li)}=0.61$ by the onset of the superwind. The production of \el{13}{C} is substantially reduced because of the weaker mixing and the model has a final \cratio ratio that is a factor of about 4 higher than in the standard case.

The EDL prescription gives similar changes in the abundances between the end of first dredge-up and the tip of the giant branch than those obtained in the standard case. This is not so surprising: we have chosen the free parameter of this prescription such that it gives roughly the same level of carbon depletion as the standard case, so we would expect the other species to behave in a similar way. However, the EDL prescription leads to significant depletion of \el{3}{He} and \el{7}{Li} at the beginning of the core helium  burning phase. By the time the star reaches the AGB, the surface \el{3}{He} abundance has dropped to just $1.04\times10^{-4}$, nearly a factor of 3 less than the standard case. This severely limits the action of thermohaline mixing along the AGB. With less \el{3}{He} available, the mixing efficiency is substantially reduced and so the drop in the \el{3}{He} along the TP-AGB is much less than in the standard case. In addition, less \el{13}{C} can be transported up from the burning shell and the model reaches the superwind phase with a very high \cratio ratio. \el{7}{Li} production suffers two-fold.  There is less \el{3}{He} available from which to produce \el{7}{Be} and the reduction in the transport efficiency means more beryllium and lithium is destroyed in the stellar interior before it can reach the safety of the convective envelope. The final surface lithium abundance is over 2 dex lower than in the standard case, with $\log_{10} \epsilon\mathrm{(^7Li)}$ being just 0.28. 

There are also uncertainties that we are unable to assess at present. Foremost among these is the suggestion that thermohaline mixing can be inhibited by the presence of rotation \citep{2008ApJ...684..626D}. Until hydrodynamic simulations of the interactions of these two mechanisms are made, this point cannot be addressed. It is also probable that other effects (e.g. the presence of magnetic fields) will also influence the degree of mixing.

\section{Discussion}

The action of thermohaline mixing during the TP-AGB does reduce the \cratio ratio, so that the ejecta will have a ratio of a few thousand rather than the $10^5$ predicted by canonical models \citep[e.g.][]{2008MNRAS.389.1828S}. While this is an improvement it still leaves the models predicting higher \cratio ratios than are observed in CEMP turn-off stars. While more evolved CEMP stars will undergo mixing of CN-cycled material on the first giant branch (either at first dredge-up or through extra mixing above the luminosity bump) which will lower their surface \cratio ratios, the turn-off objects are not evolved enough for this to have happened.  Another mechanism for reducing this ratio must clearly be sought. It is also unable to affect the \el{14}{N} abundance and so cannot account for the correlation of C- and N-enhancement observed in the CEMP stars\footnote{\citet{2009arXiv0901.4737M} suggest that the apparent C and N correlation does not hold when considering the individual CEMP subclasses. Therefore, the failure of the model to produce a C and N correlation may not be so serious.}. However, if mixing occurs to regions below the point of minimum $\mu$ due to some sort of overshooting \citep{2008ApJ...679.1541D}, it is possible that substantial enhancements of \el{13}{C} and \el{14}{N} could result.

The surprising outcome of these simulations is the high Li abundances that  can be produced. It is usually supposed that only the higher mass AGB stars which undergo hot bottom burning are able to produce Li via the Cameron-Fowler mechanism \citep{1971ApJ...164..111C}. This work shows that it is possible that {\it low-mass AGB stars could be producers of lithium-7}. We also note that the action of thermohaline mixing on the AGB is subtly different from its action of the RGB. On the RGB, it leads to a depletion of Li with the star leaving the RGB with virtually no lithium left. However, on the AGB thermohaline mixing can substantially increase the Li abundance up to values above the Spite plateau value. It is therefore possible that low-mass AGB stars have contributed to the Galaxy's Li budget. The effect of a population of low-mass, low-metallicity lithium producers on Galactic chemical evolution models should be investigated.

These models do improve the agreement of the AGB models with observations of Li in carbon-enhanced metal-poor stars. We have extracted from the Stellar Abundances for Galactic Archaeology (SAGA) database \citep{2008arXiv0806.3697S} those CEMP stars that are both C-rich and have measured Li-abundances, and also that are still close to the main sequence turn-off (because first dredge-up will significantly reduce the surface Li abundance). We select only those stars in the metallicity range $-3<$[Fe/H]$<-2$ as the models presented herein may be expected to apply only over a limited range in metallicity. In particular, at very low metallicities,  AGB stars can undergo additional mixing events not present in our models \citep[see e.g.][among many others for a discussion of these mixing events]{1990ApJ...349..580F,2008A&A...490..769C,2009MNRAS.396.1046L}. SAGA lists 5 turn-off objects with measured lithium abundances, the properties of which are displayed in Table~\ref{tab:LiCEMPs}.

One caveat should be added to the following discussion. The scenario of mass transfer from an AGB primary star on to a lower mass secondary in a binary system is expected to apply to those stars belonging to the CEMP-$s$ subclass, i.e. those stars which have [Ba/Fe]$>1$ and [Ba/Eu]$>$0.5 according to the definitions given by \citet{2005ARA&A..43..531B}. The origin of the $r/s$ subclass of CEMP stars, which have $0<$[Ba/Eu]$<0.5$, is currently unknown. It is possible that their $s$-process enrichment has come from a binary mass transfer event, in which case the models presented herein would apply, but the enrichment may have another source entirely \citep[see e.g.][for a possible alternative formation scenario]{2009PASA...26..322L}. This should be borne in mind throughout the following discussion.

\begin{table}
\begin{center}
\begin{tabular}{lccccr}
\hline
Object & [Fe/H] & [C/Fe] & $\log_{10} \epsilon\mathrm{(Li)}$ & $\log$ g & Refs. \\
\hline
CS~22964-161A & -2.41 & 1.35 & 2.09 & 3.7 & 1 \\
CS~22964-161B & -2.39 & 1.15 & 2.09 & 4.1 & 1 \\
HE~0024-2523 & -2.7 & 2.6 & 1.5 & 4.3 &  2  \\ 
CS~31080-095 & -2.85 & 2.69 & 1.73 & 4.5 &  3 \\
CS~31062-012 & -2.53 & 2.14 & $2.3$ & 4.3 & 4 \\ 
\hline
\end{tabular}
\end{center}
\caption{Properties of CEMP turn-off stars with measured Li-abundances, as extracted from the SAGA database. References: 1 -- \citet{2008ApJ...677..556T}, 2 -- \citet{2003AJ....125..875L}, 3 -- \citet{2006A&A...459..125S}, 4 -- \citet{2008ApJ...678.1351A}}
\label{tab:LiCEMPs}
\end{table}

\subsection{CS~22964-161A \& B}
\citet{2008ApJ...677..556T} reported abundances for this intriguing system. It is a double CEMP binary, with both components being turn-off objects. It is presumed that this system is actually a hierarchical triple system, in which an AGB star polluted the tight inner binary which we now observe today.

As both objects in the system have [C/Fe]=1.21 we are forced to assume that some mixing of accreted material has taken place in order to get the surface [C/Fe] low enough\footnote{Alternatively, the AGB models could be over-predicting the amount of C that is dredged-up. A simple back of the envelope calculation suggests that to reach [C/Fe] = +1 about $5\times10^{-4}$\ms\ of intershell material would have to be dredged-up, compared with around $4\times10^{-3}$\ms\ in the 1\ms\ model.}, as the three AGB models presented here all give [C/Fe]$>2$ in their ejecta. \citet{2009MNRAS.394.1051S} looked at models in which thermohaline mixing and gravitational settling are allowed to occur. In addition,  an extra turbulent mixing process might also be at work. \citet{2005ApJ...619..538R} suggested that the Spite plateau value for Li could be reconciled with the Big Bang Nucleosynthesis predictions if some unknown turbulent mixing process near the stellar surface could carry Li down to temperatures where some of it could be destroyed. The physical cause of such a process remains elusive and the prescription employed is an {\it ad hoc} one. 

\begin{figure}
\includegraphics[width=\columnwidth]{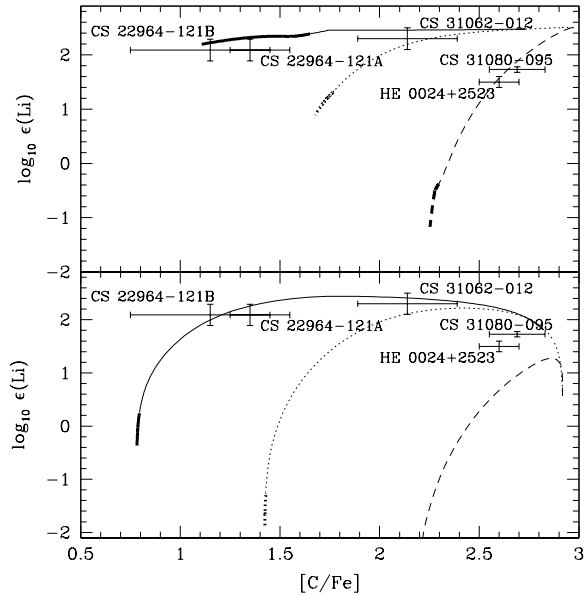}
\caption{The evolution of $\log_{10}\epsilon\mathrm{(Li)}$ with [C/Fe] when accreting material from a 1.5\ms\ companion. The cases displayed are for when 0.001\ms\ (solid line), 0.01\ms\ (dotted line) and 0.1\ms\ (dashed line) is accreted. In each case, the secondary is left with a total mass of 0.8\ms. Bold lines indicate where $\log g$ passes from 4.5 to 3.5 as the object evolves off the main sequence. The errorbars denote the locations of specific observed systems. {\bf Top panel:} The secondary is modelled including thermohaline mixing, gravitational settling and an extra turbulent process. {\bf Bottom panel:} The secondary is modelled with thermohaline mixing alone.}
\label{fig:LivCFe}
\end{figure}

Models of the secondary involving these three processes \citep[see][for details]{2009MNRAS.394.1051S} were run, accreting 0.001-0.1\ms\ of material from companions of 1, 1.5 and 2\ms, in each case producing a star of 0.8\ms. The evolution of these models through the [C/Fe]-$\log_{10}\epsilon\mathrm{(Li)}$ plane are shown in the top panel of Fig.~\ref{fig:LivCFe}. We find that we can reproduce the observed properties of CS~22964-161 if around 0.002\ms\ of material was accreted from the 1.5\ms\ companion, with star A accreting slightly more mass than star B.  In contrast, models including thermohaline mixing alone deplete far too much of their lithium to match the observations (lower panel of Fig.~\ref{fig:LivCFe}), as was noted by \citet{2009MNRAS.394.1051S}.

\subsection{HE~0024-2523}
Detailed abundance analyses for this object have been carried out by \citet{2002AJ....124..481C} and \citet{2003AJ....125..875L}. These authors report [Fe/H] for this object as being -2.62 and -2.72 respectively. Both studies agree that the surface gravity of the object is $\log g=4.3$. The bulk of the abundances for this object come from the latter work and it is from here that we quote the abundances [C/Fe]=2.6 and $\log_{10}\epsilon\mathrm{(Li)=1.5}$. The barium abundance according to \citet{2003AJ....125..875L} is [Ba/Fe]=1.46, while \citet{2002AJ....124..481C} give [Ba/Fe] = 1.76. The object is therefore $s$-process rich. Only an upper limit for europium, [Eu/Fe]$<1.16$ is given by \citet{2003AJ....125..875L} and so it cannot be ruled out that HE~0024-2523 belongs to the r/s CEMP sub-class. However, these authors do report that the star is extremely lead-rich, with [Pb/Fe] = 3.3.

HE~0024-2523 could conceivably have accreted material from a companion in a similar mass range to the models presented here. The [C/Fe] and $\log_{10} \epsilon\mathrm{(Li)}$ values of the models are in the region of the values observed in this object. We would have to assume that no mixing of the accreted material took place, as any mixing would reduce the C- and Li abundances too much. 

\subsection{CS~31080-095}
\citet{2006A&A...459..125S} have performed a detailed abundance analysis of this object. They find [Fe/H] = -2.85, [C/Fe] = 2.69 and $\log_{10}\epsilon\mathrm{(Li)}=1.73$. A moderate barium enhancement of [Ba/Fe] = 0.77 is reported, but this is the heaviest element for which they give an abundance.

CS~31080-095 could conceivably come from a similar mass range to the models presented here. The C and Li abundances of this object lie between those of our 1 and 1.5\ms\ models, so we would expect that accreting material from a companion in this mass range would give appropriate abundances, provided the accreted material did not mix into the secondary.

\subsection{CS~31062-012}

CS~31062-012 has been studied by \citet{1997ApJ...488..350N}, \citet{2001ApJ...561..346A}, \citet{2002PASJ...54..933A}, \citet{2002ApJ...580.1149A} and \citet{2008ApJ...678.1351A}. These authors give [Fe/H] in the range -2.53 to -2.74, with [C/Fe] in the range 2.1 to 2.15. \citet{2008ApJ...678.1351A} give the lithium abundance as $\log_{10}\epsilon\mathrm{(Li)}=2.3$, but an earlier study by \citet{2005A&A...442..961C} gives a lower value of $\log_{10}\epsilon\mathrm{(Li)}=1.973$. The object is barium-rich, with [Ba/Fe] around 2, and somewhat enriched in europium with [Eu/Fe] = 1.4 being reported by both \citet{1997ApJ...488..350N} and \citet{2002ApJ...580.1149A}. This would place CS~31062-012 in the CEMP-$s$ subclass.

We find we are able to model this system using ejecta from a 1.5\ms\ companion. If around 0.002\ms\ of material is accreted on to the secondary, and both thermohaline mixing and gravitational settling are included (but not the additional turbulent process), then the properties of the system can be reproduced. Dilution of the small quantity of material accreted allows the carbon abundance to drop to the requisite level, while the gravitational settling creates a mean molecular weight barrier that prevents the mixing going too deep, allowing a substantial quantity of lithium to survive. Inclusion of the extra turbulent mixing process of \citet{2005ApJ...619..538R} leads to a degree of carbon depletion that is too great. 

It is rather unsatisfactory that for some objects we seem to require some degree of mixing of the accreted material and for others we require that no mixing of this material takes place. Similar conclusions were reached by \citet{2009MNRAS.394.1051S} when examining the trends of various light elements in CEMP stars. The physical reason why some stars mix their accreted material and others do not remains unknown. It is likely that the interaction of these mixing processes with other physical processes not included here (e.g. rotation, magnetic fields) is responsible.

\subsection{Systems with upper limits on lithium}
In addition to the above systems which have measurements of the lithium abundances, SAGA lists five systems for which an upper limit for lithium has been determined. These systems are: CS~29497-030, CS~29526-110, SDSS~0924+40, SDSS~1707+58 and SDSS~2047+00. The salient details of these models are given in Table~\ref{tab:LiLimitCEMPs}. The high upper limits derived for the latter four systems do not severely restrict which models can be applied. For example, a model involving accretion of material from a 1\ms\ AGB star, with no mixing of the accreted material into the secondary would give surface abundances broadly consistent with the observations of all but SDSS~0924+40. However, accretion of ejecta from a more massive AGB companion would also fit the data providing some mixing of the accreted material took place in order to lower the surface carbon abundance of the secondary. Without better constrains on the lithium abundance, it is not possible to determine which of these scenarios actually took place.

\begin{table}
\begin{center}
\begin{tabular}{lccccr}
\hline
Object & [Fe/H] & [C/Fe] & $\log_{10} \epsilon\mathrm{(Li)}$ & $\log$ g & Refs. \\
\hline
CS~29497-030 & -2.7 & 2.38 & $<$1.16 & 3.5 & 1 \\
CS~29526-110 & -2.1 & 2.07 & $<2.3$ & 4.1 & 2 \\
SDSS~0924+40 & -2.51 & 2.72 & $<2$ & 4.0 & 2 \\
SDSS~1707+58 & -2.52 & 2.1 & $<2.5$ & 4.2 & 2 \\
SDSS~2047+00 & -2.05 & 2.3 & $<2.3$ & 4.5 & 2 \\
\hline
\end{tabular}
\end{center}
\caption{Properties of CEMP turn-off stars with upper limits on the Li abundance, as extracted from the SAGA database. References: 1 -- \citet{2004A&A...413.1073S},  2 -- \citet{2008ApJ...678.1351A}.}
\label{tab:LiLimitCEMPs}
\end{table}

The systems CS~29497-030 and SDSS~0924+40 are more restrictive. The high [C/Fe] of SDSS~0924+40 rules out accretion from a 1\ms\ companion because this model does not produce enough carbon. We can also rule out the case of accretion of material from a 1.5\ms\ model. While this model produces roughly the correct amount of [C/Fe], its Li abundance is about 0.5 dex above the observed upper limit. Accretion from a more massive companion with some mixing of the accreted material is a viable scenario. For CS~29497-030, accretion from a 1\ms\ companion is consistent with the observed abundances, provided no mixing of the accreted material takes place. However, it is also possible the companion was more massive and that mixing of the accreted material happened.

\section{Conclusions}

We have investigated the effect that thermohaline mixing has on the abundances of low-mass, low-metallicity AGB stars. We find that enough \el{3}{He} remains after the first giant branch that thermohaline mixing can still take place on the AGB. However, the effect is only felt during the very first thermal pulses with dredge-up. Thermohaline mixing can lead to substantial production of \el{7}{Li} -- even up to values above the Spite plateau value. Thus it is possible to reconcile C- and Li-rich metal-poor stars with having come from a binary mass transfer scenario. We demonstrate that the Li-enrichments of five known Li-rich, turn-off CEMP stars can be explained using the abundances of our models. Thermohaline mixing does not reduce the \cratio ratio sufficiently to reconcile the AGB models with observations of CEMP stars at the Main Sequence turn-off, nor does it raise the surface nitrogen abundance sufficiently.
The possibility that low-mass, low-metallicity stars could be producers of lithium is intriguing and their role in Galactic chemical evolution should be investigated.

Many uncertainties about the nature of thermohaline mixing still remain and we have shown that the Li abundance is particularly sensitive to the nature of the mixing. There is also the question of how other physical processes (e.g. rotation) interact with thermohaline mixing and the consequences of this remain to be explored. 

\section{Acknowledgements}
The anonymous referee is thanked for her/his useful comments that have helped to improve the clarity of this manuscript. The author thanks Evert Glebbeek for reading the manuscript prior to submission. RJS is funded by the Australian Research Council's Discovery Projects scheme under grant DP0879472. This work was supported by the NCI National Facility at the ANU.

\bibliography{../../../masterbibliography}

\label{lastpage}

\end{document}